\documentclass[conference,a4paper]{IEEEtran}
\IEEEoverridecommandlockouts
\usepackage[noadjust]{cite}
\usepackage{amsmath,amssymb,amsfonts}
\usepackage{algorithmic}
\usepackage{graphicx}
\usepackage{textcomp}
\usepackage{xcolor}

\usepackage{xcolor}

\usepackage{tikz}

\usetikzlibrary{shapes,arrows,positioning,patterns}
\tikzstyle{data} = [draw, rectangle, text width=6cm, text centered, minimum height=0.8cm, node distance=1.5cm ]
\tikzstyle{block} = [draw, rectangle, text width=6cm, text centered, minimum height=0.8cm, node distance=1.5cm, fill=blue!20]
\tikzstyle{speechblock} = [draw, rectangle, text width=6cm, text centered, minimum height=0.8cm, node distance=1.5cm, pattern=north west lines, pattern color=blue!20]
\tikzstyle{audioblock} = [draw, rectangle, text width=4cm, text centered, minimum height=0.8cm, node distance=1.5cm, pattern=dots, pattern color=blue!20]
\tikzstyle{line} = [draw, -latex']
\tikzstyle{container} = [draw, rectangle, inner sep=0.3cm]

\def\BibTeX{{\rm B\kern-.05em{\sc i\kern-.025em b}\kern-.08em
    T\kern-.1667em\lower.7ex\hbox{E}\kern-.125emX}}
\begin{document}

\title{ViSQOL v3: An Open Source Production Ready Objective Speech and Audio Metric \\
\thanks{Available at https://github.com/google/visqol}
}

\author{\IEEEauthorblockN{Michael Chinen, Felicia S. C. Lim, and Jan Skoglund}
\IEEEauthorblockA{\textit{Chrome Media Audio} \\
\textit{Google LLC}\\
San Francisco, USA \\
{mchinen,flim,jks}@google.com}
\and
\IEEEauthorblockN{Nikita Gureev}
\IEEEauthorblockA{\textit{Hangouts Meet} \\
\textit{Google LLC}\\
Stockholm, Sweden \\
gureev@google.com      }
\and
\IEEEauthorblockN{Feargus O'Gorman and Andrew Hines}
\IEEEauthorblockA{\textit{School of Computer Science} \\
\textit{University College of Dublin}\\
Dublin, Ireland \\
feargusog@gmail.com, andrew.hines@ucd.ie}
}

\IEEEpubid{\makebox[\columnwidth]{978-1-7281-5965-2/20/\$31.00 
\copyright 2020 IEEE \hfill} 
\hspace{\columnsep}\makebox[\columnwidth]{ }}

\maketitle

\begin{abstract}
  Estimation of perceptual quality in audio and speech is possible using a  variety of methods. The combined v3 release of ViSQOL and ViSQOLAudio (for speech and audio, respectively,) provides improvements upon previous versions, in terms of both design and usage. As an open source C++ library or binary with permissive licensing, ViSQOL can now be deployed beyond the research context into production usage. The feedback from internal production teams at Google has helped to improve this new release, and serves to show cases where it is most applicable, as well as to highlight limitations. The new model is benchmarked against real-world data for evaluation purposes. The trends and direction of future work is discussed.
\end{abstract}

\begin{tikzpicture}[overlay, remember picture]
\path (current page.north) node (anchor) {};
\node [below=of anchor] {%
2020 Twelfth International Conference on Quality of Multimedia Experience (QoMEX)};
\end{tikzpicture}

\section{Introduction}
There are numerous objective metrics available, i.e, metrics obtained by measurements on the audio signal, to assess the quality of recorded audio clips. Examples of physical measures include signal-to-noise ratio (SNR), total harmonic distortion (THD), and spectral (magnitude) distortion. When estimating perceived quality, PESQ \cite{rix2001perceptual, pesq} and POLQA \cite{beerends2013perceptual, polqa3} have become standards for speech, and in practice also for general audio, despite being originally designed to target only speech quality. There are other notable examples, e.g., PEAQ \cite{thiede2000peaq} and PEMO-Q \cite{huber2006pemo}. Most of these metrics require commercial licenses. ViSQOL \cite{hines2012visqol} and ViSQOLAudio \cite{hines2015visqolaudio} (referred to collectively as ViSQOL below), are freely available alternatives for speech and audio.  These metrics are continually being expanded to cover additional domains.  For example the work on AMBIQUAL \cite{narbutt2018ambiqual} extends the same principles used in ViSQOLAudio into the ambisonics domain.

Advancements in speech and audio processing, such as denoising and compression, propel the need for improvements in quality estimation. For example, speech and audio codecs reach lower and lower useful bitrates.  As such, it may be worthwhile to analyze the performance of ViSQOL for this extended domain. Furthermore, there have been a number of deep neural network (DNN) generative models that recreate the waveform by sampling from a distribution of learned parameters.  One example is the WaveNet-based low bitrate coder \cite{kleijn2018wavenet}, which is generative in nature. There are other DNN-based generative models, including SampleRNN \cite{klejsa2019high} and WaveGlow \cite{prenger2019waveglow}, with promising results that suggest that this trend will continue. These generative models typically do not lend themselves to being analyzed well by existing full reference speech quality metrics. While the work described in this paper does not propose a solution to the generative problem, the limitations of the current model should be acknowledged to encourage development of solutions.

ViSQOL was originally designed with a polynomial mapping of the neurogram similarity index measure (NSIM)~\cite{Hines2012} to MOS, and ViSQOLAudio was extended to use a model trained for support vector regression. Since then, deep neural network models have emerged and been applied to speech quality models~\cite{avila2019non, gamper2019intrusive}. Such approaches are promising and potentially can resolve some of the issues that the current architectures cannot. While such new directions are clearly interesting and warrant further investigation, they are rapidly evolving.

we present a new version of ViSQOL, v3, which contains incremental improvements to the existing framework based on real-world feedback, rather than fundamental changes such as end-to-end DNN modeling.  Since ViSQOL has been presented and benchmarked in a large number of experiments that have validated its application to a number of use cases \cite{hines2012visqol, hines2013robustness, hines2015measuring, hines2015visqolaudio, sloan2016bitrate, sloan2017objective} we consider it relatively well analyzed for the known datasets, which tend to be smaller and relatively homogeneous.  We instead turn our attention to the data and types of problems encountered ``in the wild'' at Google teams that were independent of ViSQOL development, and the iterative improvements that have come from this analysis.  Adapting it to these cases has yielded various improvements to usability and performance, along with feedback and insights about the design of future systems for estimating perceptual quality.  Since the nature these improvements fill the 'blind spots' in the datasets, they are not expected to improve its results on these datasets.  Until there is the creation of more diverse subjective score datasets, real-world validation seems to be a reasonable compromise.

Additionally and alongside improving the quality of MOS estimation from real-world data, we are concerned with how to make ViSQOL more useful to the community from a practical tooling perspective.  Even though ViSQOL was available through a MATLAB implementation, there were still unnecessary hurdles to use it in certain cases, e.g. production and continuous integration testing, (which may need to run on a server), or may not have MATLAB licenses available.  As a result, we chose to re-implement it in C++ because it is a widely available and extensible language that can be wrapped in other languages.  We decided to put the code on GitHub for ease of access and contribution.

This paper is structured as follows: in section \ref{sec:case_studies}, a case study of the findings and challenges encountered when integrating ViSQOL into various Google projects. In section \ref{sec:design}, we present the general design and algorithmic improvements that are in the new version. Then in section \ref{sec:discussion}, the improvements with respect to the case studies are discussed. Finally, we summarize in a concluding section \ref{sec:conclusion}.

\section{Case Studies and User Feedback}
\label{sec:case_studies}
This version of ViSQOL is the result of the integration process of ViSQOL, using real production and integration testing cases at Google.
The case studies described in this section were initiated by individual teams that were independent of prior ViSQOL development. They typically consulted with a ViSQOL developer to verify appropriate usage, or read the documentation and integrated ViSQOL on their own.
\subsection{Hangouts Meet}
The Meet team has been successfully using ViSQOL for assessing audio quality in Hangouts Meet. Hangouts Meet is a video communication service that uses WebRTC \cite{WebRTC} for transmitting audio. Meet uses a testbed that is able to reliably replicate adverse network conditions to assess the quality of audio during the call. For this use case they have 48~kHz-sampled reference and degraded audio samples and use ViSQOLAudio for calculating the results.

\begin{figure}[htbp]
\begin{center}
  \includegraphics[width=.5\textwidth]{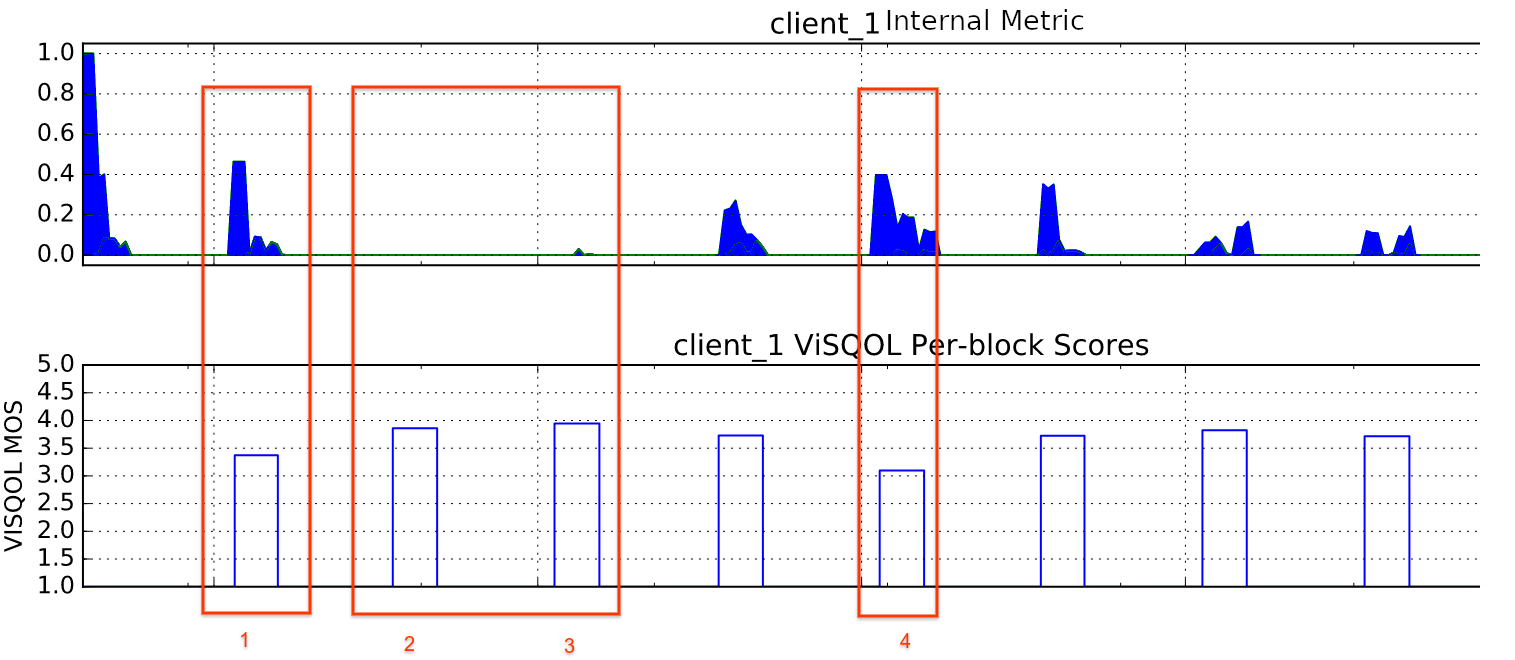} \\
  \caption{Hangouts Meet's internal no-reference metric has components to detect audio degradations. ViSQOL successfully detected these degradations in audio that contained them (blocks 1 and 4), while in the audio blocks that were not affected the scores from ViSQOL were higher (blocks 2 and 3).}
\label{fig:meet_boxes}
\end{center}
\end{figure}

In order to ensure that ViSQOL works reliably for this use case, it was compared to an internal no-reference audio quality metric that is based on technical metrics of a WebRTC-based receiver. The metric is on a scale from 0 to 1, with lower scores being better. ViSQOL's MOS is able to correlate to this metric, as seen in Figure \ref{fig:meet_boxes}.

\begin{figure}[htbp]
\begin{center}
  \includegraphics[width=.5\textwidth]{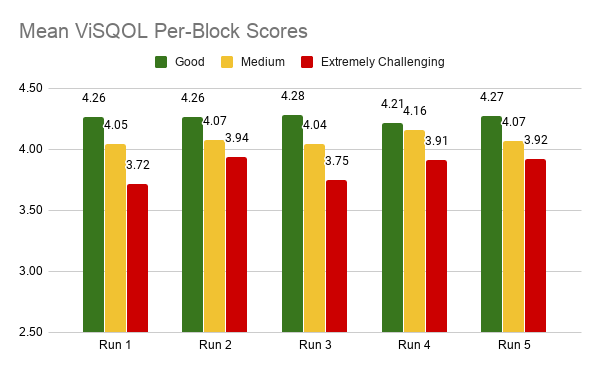} \\
  \caption{Comparison between mean ViSQOL MOS and network degradations. Some of the calls were run with good network conditions (green), some were simulating average network conditions, where the product should still perform well (yellow), while others were simulating extremely challenging network conditions, where it is expected to for issues to appear (red). }
\label{fig:meet_per_block_mean}
\end{center}
\end{figure}

In this use case, Meet developers were mostly interested in the sensitivity of ViSQOL to audio degradations from network impairments. In Figure \ref{fig:meet_per_block_mean} there is a comparison between mean ViSQOL scores during a call that shows that the metric is sensitive to how audio quality changed from a good network conditions scenario with scores ranging from 4.21 to 4.28, to a medium impaired scenario with scores ranging from 4.04 to 4.16, to finally an extremely challenging network scenario with scores from 3.72 to 3.94. Although the exact network conditions can not be shared, here good network conditions indicated that the connection should allow for both video and audio to be near perfect in the call, medium conditions indicate that the call might have issues, but the audio should continue to be good, while in extremely challenging conditions we expect to see both video and audio perceptually degraded, but the call would still go through.

\begin{figure}[htbp]
\begin{center}
  \includegraphics[width=.5\textwidth]{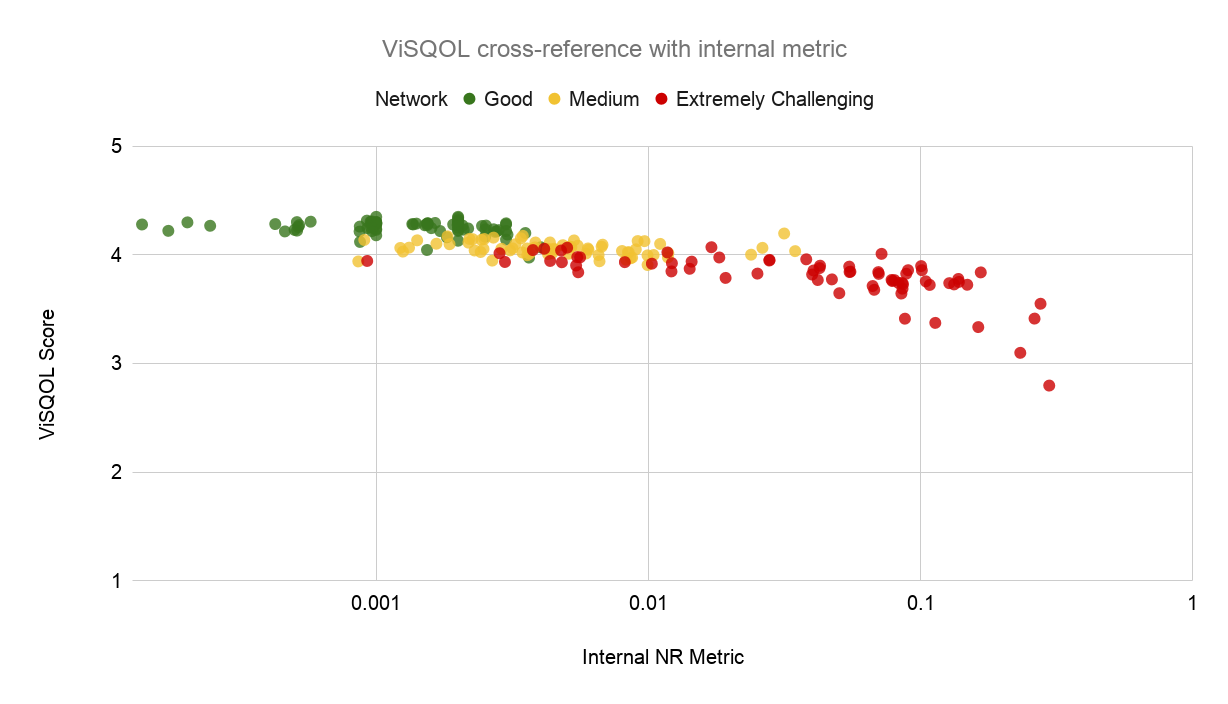} \\
  \caption{Scatter plot of ViSQOL MOS versus a no-reference internal metric. Each point represents a call.}
\label{fig:meet_cross_reference}
\end{center}
\end{figure}

In order to ensure that ViSQOL performs reliably, several hundreds of calls were collected from the testbed.  The mean values obtained from ViSQOL and the internal metric from these calls were plotted in Figure \ref{fig:meet_cross_reference}. The results were reliably reproduced. Following the positive results from this investigation, ViSQOL is currently one of the main objective audio quality metrics deployed by the Hangouts Meet product team at Google.

\subsection{Opus Codec}
Google contributes to the development of the Opus codec. ViSQOL and POLQA were used to benchmark the quality of the Opus coder for both speech and music at various bitrates and computational complexities. In previous studies ViSQOLAudio has been shown to perform reasonably on low bitrate audio \cite{hines2015visqolaudio}. However, ViSQOL's speech mode did not specifically target the low bitrate case. Additionally, recent advancements in Opus have pushed the lower bound of the range of bitrates further downwards for a given bandwidth since the time ViSQOL was introduced. For example, Opus 1.3 can produce a wideband signal at 9 kbps, whereas the TCDAudio14~\cite{hines2014perceived}, CoreSV14~\cite{coresv2014listening}, and AACvOpus15~\cite{sloan2017objective} datasets that ViSQOL's support vector regression was trained on have bitrates that only go as low as 24~kbps.

\begin{figure}[htbp]
\begin{center}
  \includegraphics[width=.5\textwidth]{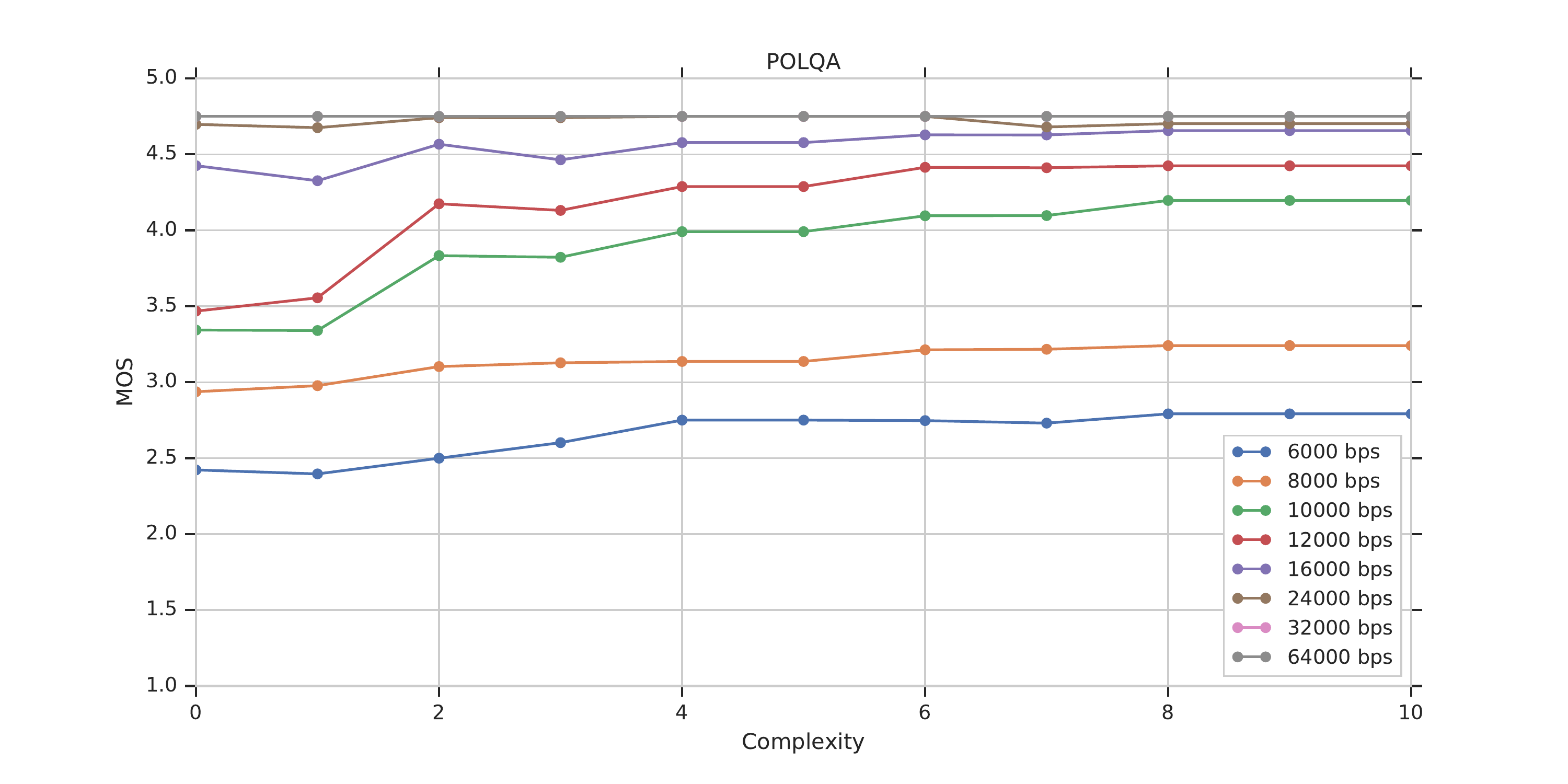} \\
  \includegraphics[width=.5\textwidth]{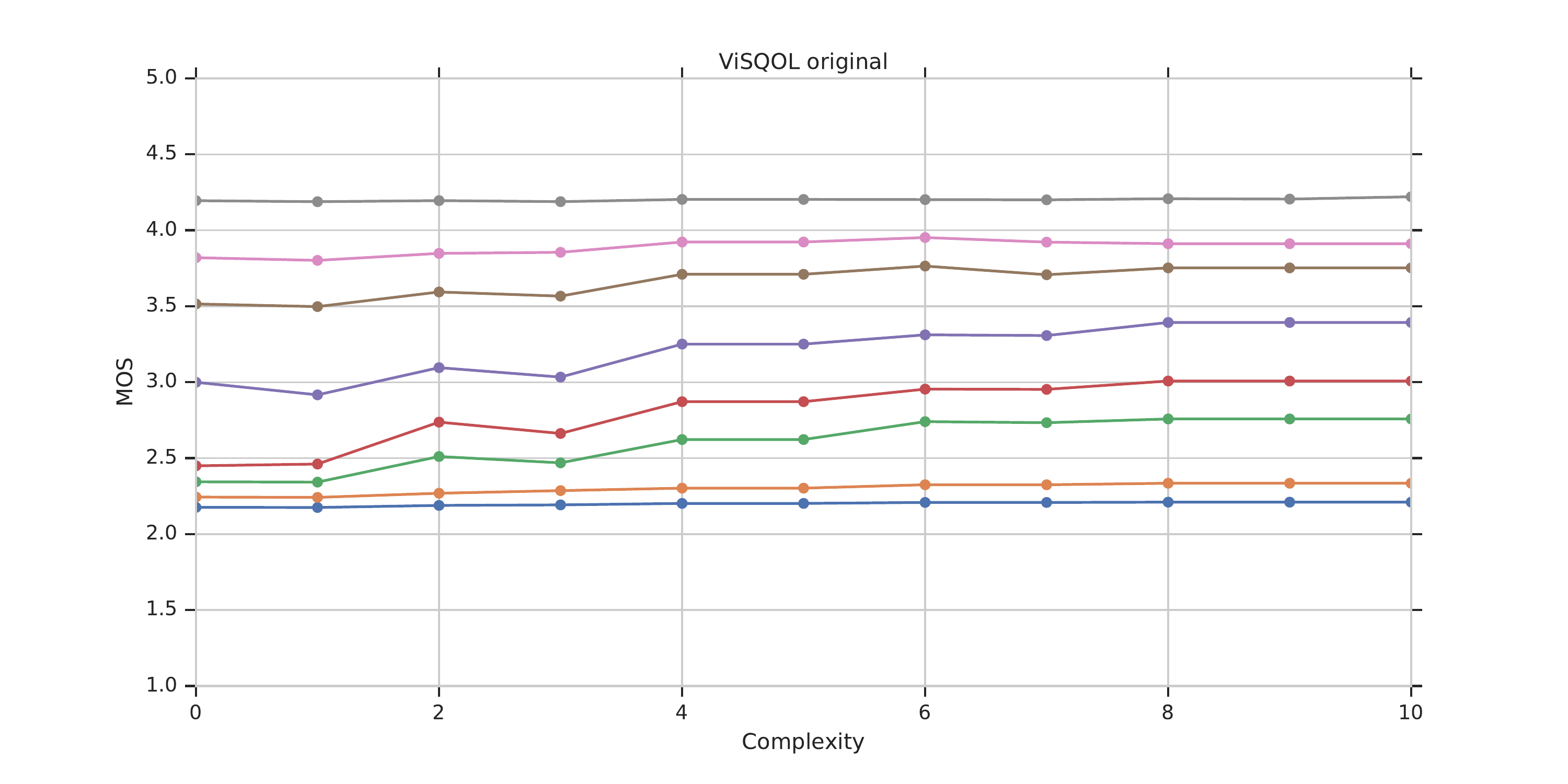} \\
  \includegraphics[width=.5\textwidth]{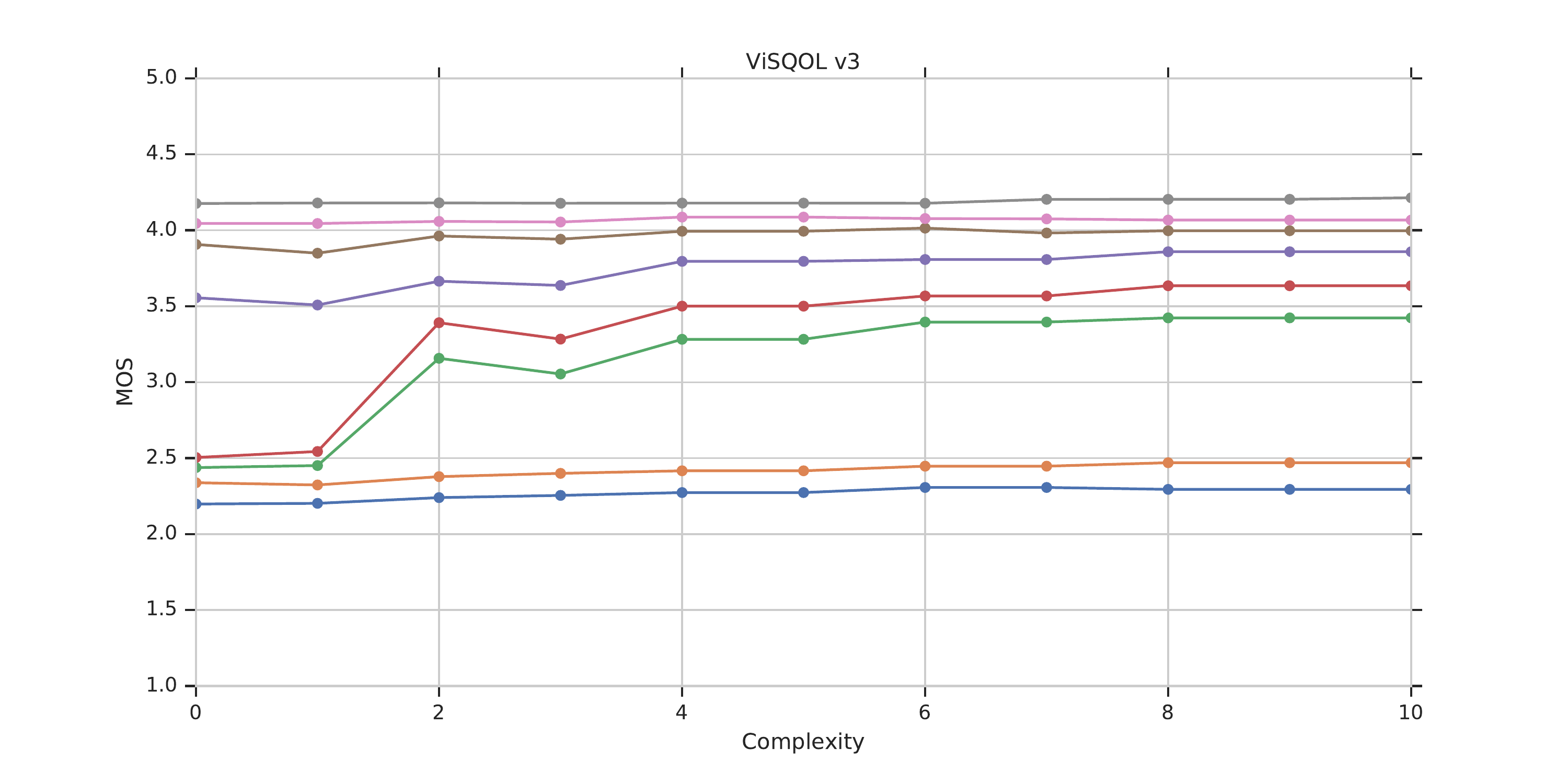} \\
  \caption{Estimated MOS for varying bitrates (6-64 kbps) and complexity settings on Opus-encoded speech.}
\label{fig:opus_speech}
\end{center}
\end{figure}

POLQA and the original version of ViSQOL in speech mode display similar trends that are consistent with expectations with respect to the bitrate and complexity settings. The differences in the \textit{lower} bitrates are more pronounced according to POLQA. The differences in \textit{higher} bitrates are more pronounced according to ViSQOL. Although subjective scores were not available, the developers expected that MOS should be less sensitive to changes in higher bitrates, giving POLQA a better match. After the improvements described in section 3, ViSQOL v3 MOS was a closer match to the expectation as can be seen in Figure \ref{fig:opus_speech}.

\begin{figure}[htbp]
\begin{center}  
  \includegraphics[width=.5\textwidth]{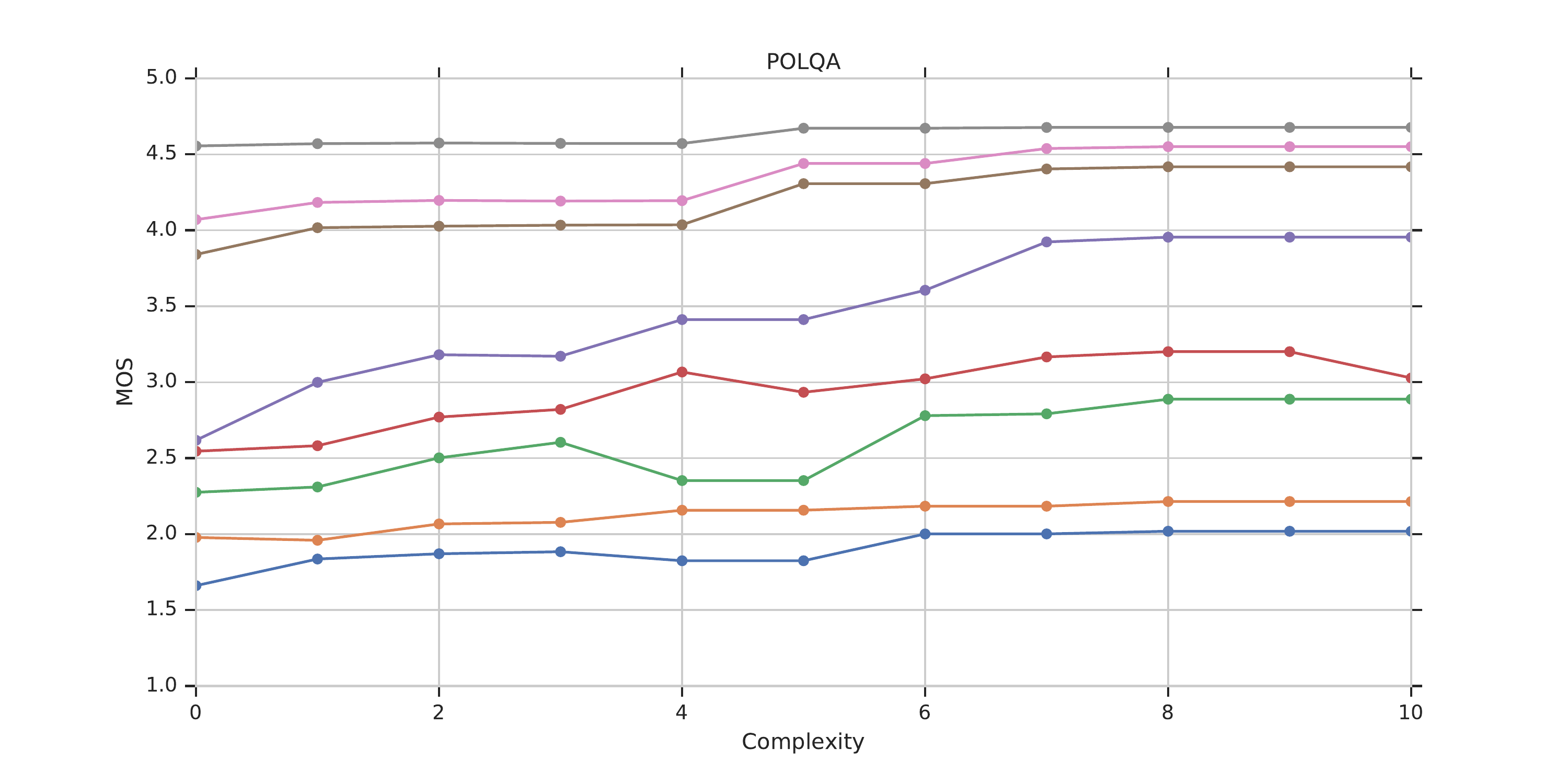} \\
  \includegraphics[width=.5\textwidth]{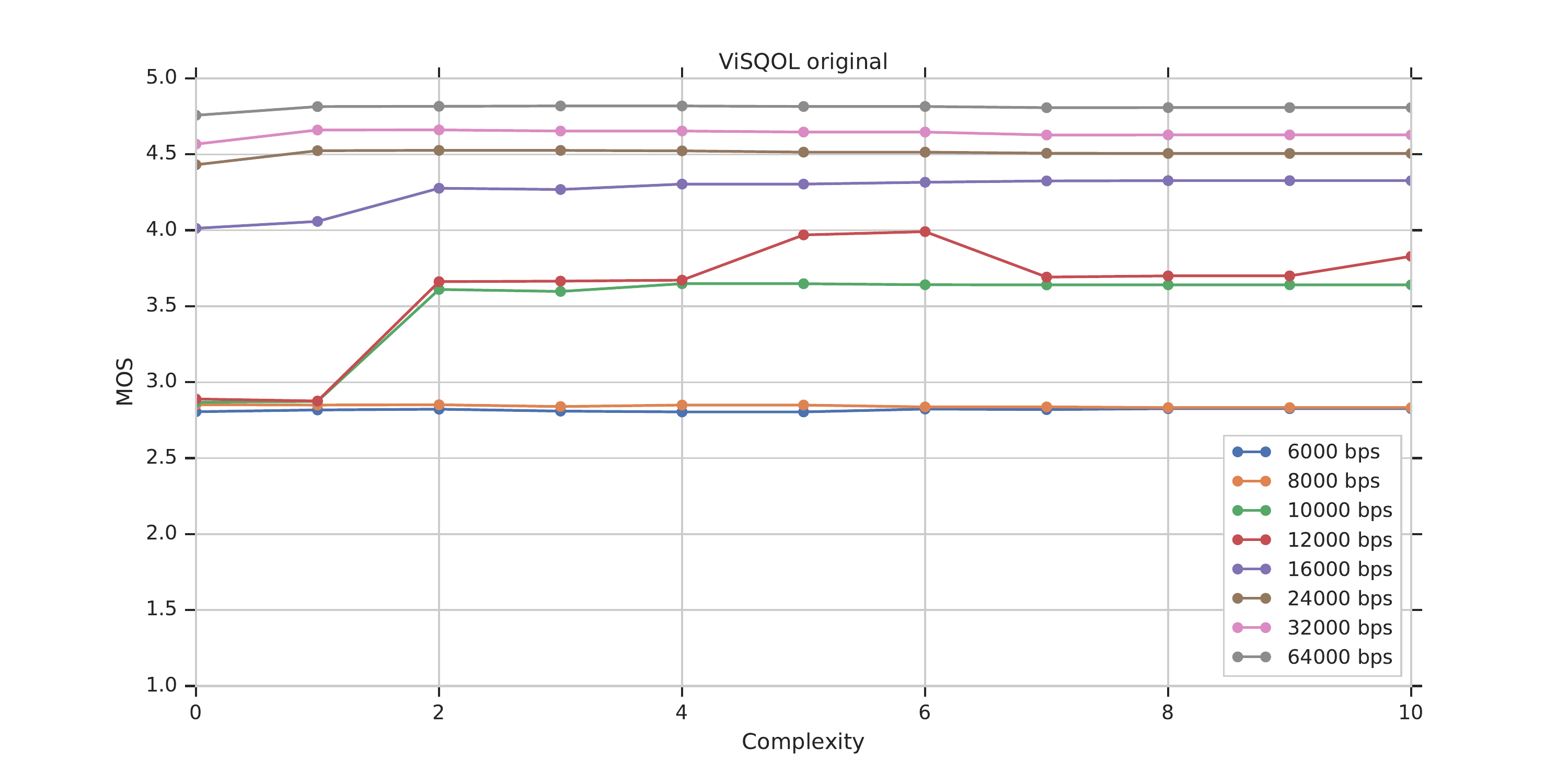} \\
  \includegraphics[width=.5\textwidth]{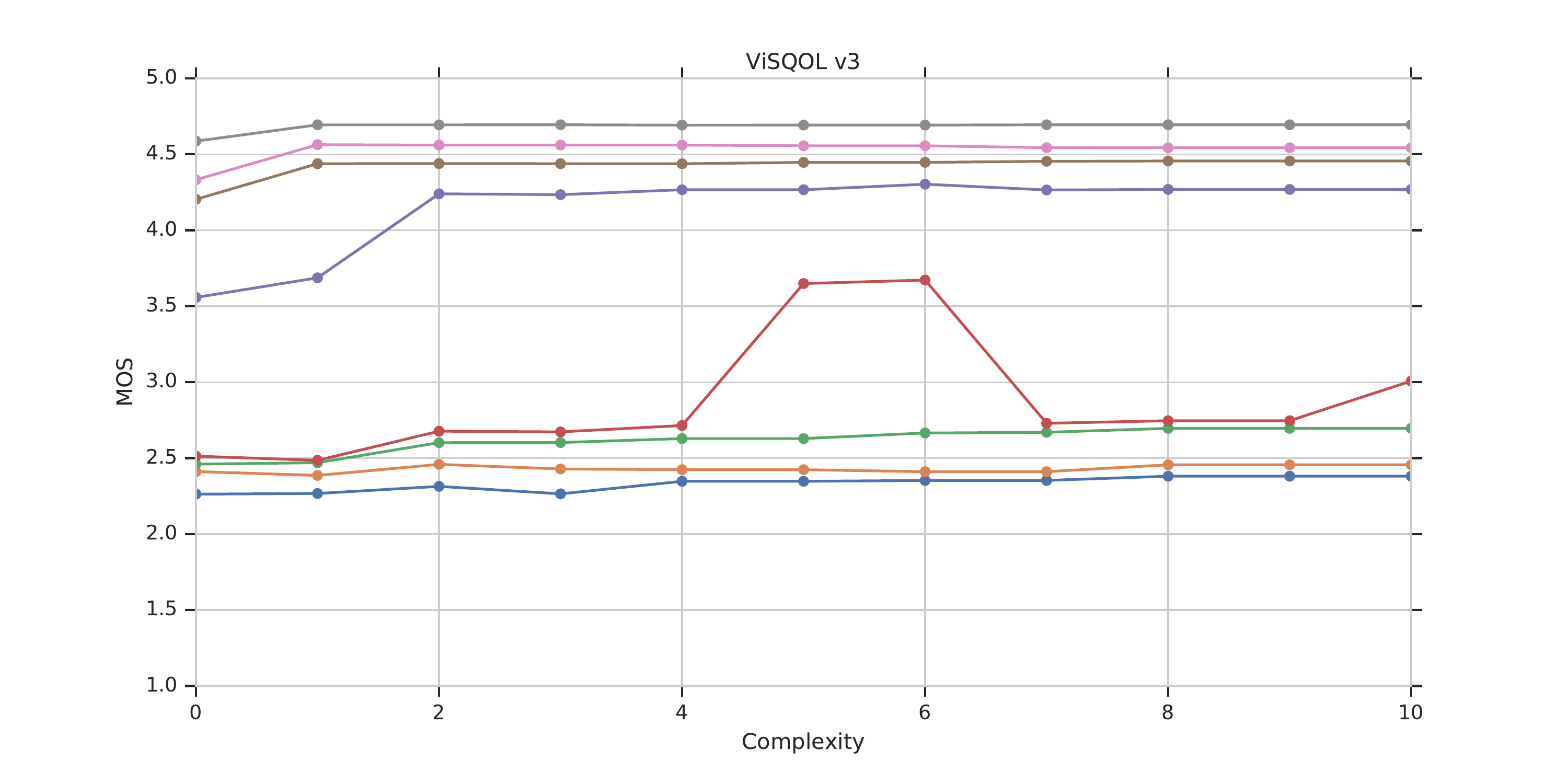} \\
  \caption{Estimated MOS for varying bitrates and complexity settings on Opus-encoded music (legend as per Fig.~\ref{fig:opus_speech}). The bitrates follow the same key as Figure \ref{fig:opus_speech}.  The bump at complexity 5, 6, and 10 for 12 kbps is related to Opus deciding to use a 12 kHz bandwidth for some fraction of the files instead of the 8 kHz bandwidth it used for complexities 2-4 and 7-9.}
\label{fig:opus_music}
\end{center}
\end{figure}

For musical examples, the developers found that both metrics display similar trends with respect to bitrates. However, POLQA shows higher discrimination between 6-8 kbps, 10-12 kbps and 16-24 kbps. ViSQOL is able to discriminate between the different bitrates with monotonic behavior, but one point of concern is that this results in ViSQOLAudio being relatively insensitive to differences in complexity settings. In light of this, we would not recommend using ViSQOL for automated regression tests without retraining the model.  The improvements made in section 3 slightly ameliorate these issues, as can be seen in Figure \ref{fig:opus_music}.  On the other hand, ViSQOL identified a spurious bandwidth `bump' at 12 kbps for the 5 and 6 complexity settings (which was perceived as higher quality in informal listening), where POLQA did not.

Lastly, ViSQOL was used to analyze the results for both clean and noisy references.  This is not a case ViSQOL was designed for, as it presumes a clean reference, similar to PESQ~\cite{pesq} and POLQA~\cite{polqa3}. However, it was found to perform in a similar fashion to the clean cases for both speech and audio in the noisy cases.

It was concluded that ViSQOL could be used for regression testing for speech. However, formal listening tests would be desirable for two reasons: to better interpret the differences between POLQA and ViSQOLAudio, and to allow training a model that represented the low bitrate ranges.

\subsection{Other Findings}
\label{subsec:otherfindings}
A number of other teams have also adapted ViSQOL for their products. In the majority of cases, their use case vaguely resembles the training data (e.g. wideband speech network degradations or music coding), but often has marked differences. For example, one team chose to analyze the network loop with a digital and analog interface, requiring a rig to be built for continuous automated testing. Typically these teams also had access to PESQ, POLQA or subjective scores for their cases and wanted to evaluate the accuracy of ViSQOL measurements as well as identify limitations.
A frequent issue was related to the duration and segmentation of the audio that would be used with ViSQOL when used in an automated framework. While ViSQOL in speech mode has a voice activity detector, it was found that ViSQOLAudio would perform poorly for segments where the reference was silent, because of either the averaging effects, or because of the lack of log-scale thresholding which was overly sensitive to small absolute differences in ambient noise levels. To resolve the averaging effects, it was recommended to extract segments of audio of 3 to 10 seconds where there was known activity. A solution to the thresholding issues is discussed in the next section.

\section{Design and improvements}
\label{sec:design}
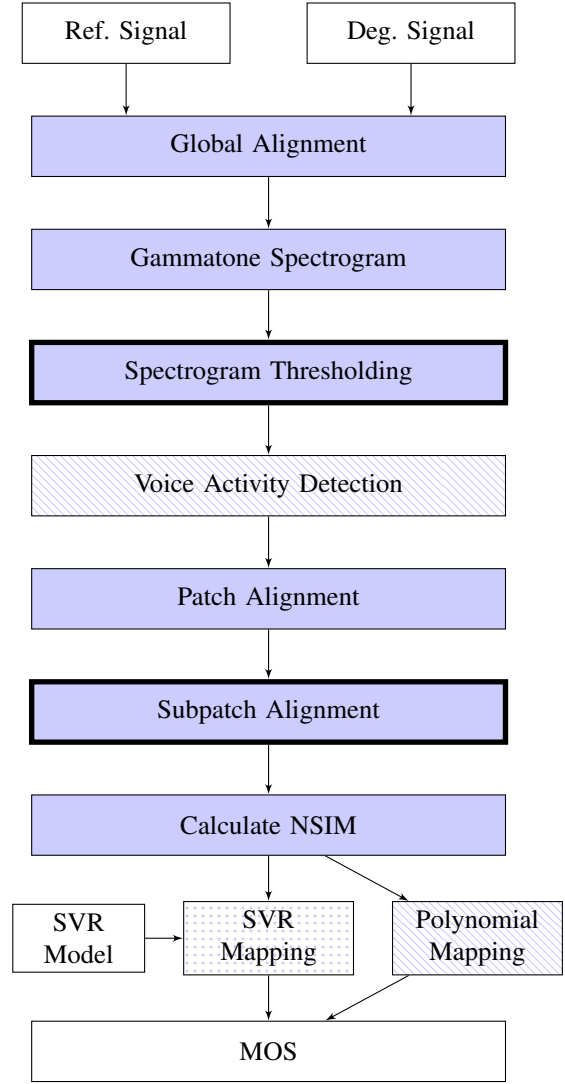
\begin{figure}[htbp]

\begin{center}
\begin{tikzpicture}
  \node [data, name=ref, text width = 2.5cm] {Ref. Signal};
  \node [data, right = 1cm of ref, name=deg, text width = 2.5cm] {Deg. Signal};
  \path (ref) -- (deg) coordinate[midway] (aux);
  \node [block, below of = aux, name=galign] {Global Alignment};
  \node [block, below of = galign, name=spec] {Gammatone Spectrogram};
  \node [block, line width = 2pt, below of = spec, name=thresh] {Spectrogram Thresholding};
  \node [speechblock, below of = thresh, name=vad] {Voice Activity Detection};
  \node [block, below of = vad, name=patch] {Patch Alignment};
  \node [block, line width = 2pt, below of = patch, name=subpatch] {Subpatch Alignment};
  \node [block, below of = subpatch, name=nsim] {Calculate NSIM};
  \node [audioblock, below of = nsim, name=mapsvr, text width = 2cm] {SVR Mapping};
  \node [data, left = .5cm of mapsvr, name=svrmodel, text width = 1.5cm] {SVR Model};
  \node [data, below of = mapsvr, name=mos] {MOS};
  \node [speechblock, below of = nsim, above of = mos, right = .5 cm of mapsvr, name=mappoly, text width = 2cm] {Polynomial Mapping};
  \path [line] (ref) -- (ref|-galign.north);
  \path [line] (deg) -- (deg|-galign.north);
  \path [line] (galign) -- (spec);
  \path [line] (spec) -- (thresh);
  \path [line] (thresh) -- (vad);
  \path [line] (vad) -- (patch);  
  \path [line] (patch) -- (subpatch);
  \path [line] (subpatch) -- (nsim);
  \path [line] (nsim) -- (mapsvr);
  \path [line] (nsim) -- (mappoly);
  \path [line] (svrmodel) -- (mapsvr);
  \path [line] (mappoly) -- (mos);
  \path [line] (mapsvr) -- (mos);  
\end{tikzpicture}
\end{center}
\caption{System Diagram. The inputs and outputs have white fill, and the processing components have blue fill. New components have thick edges. The dashed and dotted fill represent speech-only and audio-only components, respectively.}
\label{systemFig}
\end{figure}

This section summarizes the previous version and describes the changes made to the new version. Figure \ref{systemFig} shows the overall program flow and highlights the new components of the system that are referred to in the subsections.

\subsection{General Design}
The ViSQOL algorithms described in \cite{hines2012visqol} and \cite{hines2015visqolaudio} share many components by design, such as the gammatone spectrogram and NSIM calculation. It then seems reasonable that the common components be shared and developed together. The differences between the two algorithms are related to differences in the characteristics of speech and music. For example, the use of voice activity detection (VAD) for speech, and analysis of the higher bands (up to 24~kHz) for general audio/music. The common components of both speech and audio systems include creating a gammatone spectrogram using equivalent rectangular bandwidth (ERB) filters, creation of patches on the order of a half-second, aligning them, computing the NSIM from the aligned patches, and then mapping the NSIM values to MOS.

There were minor changes to some of these components because of practical reasons, such as modifying dependencies, or fixing issues found in case studies or test failures. For example, the VAD implementation uses a simple energy-based VAD, which should be sufficient given the requirement of clean references. As another example, window sizes were updated to be 80 ms with a hop of 20 ms after discovering an issue with the windowing of previous versions.

\subsection{C++ Library and Binary}
To make ViSQOL more available, we uncoupled the dependency on MATLAB by implementing a C++ version with only open source dependencies. The new version, v3, is available as a binary or as a library. The codebase was made available on GitHub because we wish for it to be easy to use by the public, and to invite external contributions.

  The majority of users were binary users, but some had requirements for finer control. For this purpose we designed a library with protobuf support and error checking, which the binary depends on.  This library would also be useful for a user that wishes to wrap the functions in a different language, such as with python bindings.
  
  There were several changes to the input and output. Verbose output has also changed to include the average NSIM values per frequency band and mean NSIM per frame. Because ViSQOL is continuously changing to adapt to new problems, a conformance version number is included in the output. Whenever the MOS changes for known files, the conformance number will be incremented. Lastly, batch processing via comma-separated value (csv) files are also supported.
  
  A number of Google-related projects were used to build this version. The application binary was implemented using the Abseil C++ application framework~\cite{abseil}. The Google Test C++ testing framework~\cite{googletest} was integrated and various tests were implemented to ensure correctness, detect regressions, and increase stability for edge cases. 23 test classes with multiple tests were implemented. These include not only unit tests, but also a test to check the conformance of the current version to known scores. The Bazel framework~\cite{bazel} was used to handle building and dependency fetching, as well as test development.

\subsection{Fine-scaled Time Alignment}
Although the previous versions of ViSQOL did two levels of alignment (global and patch), there were still issues with the patch alignment due to the spectrogram frames being misaligned at a fine scale. To address this, we implemented an additional alignment step that offsets by the lag found in a cross correlation step on the time-domain regions that corresponds to the aligned patches as described in~\cite{issc2019}. Next, the gammatone spectrogram is recomputed for sample-aligned patch audio and the NSIM score is taken.

\subsection{Silence Thresholds}
To deal with problem of log-scale amplitudes discussed in \ref{subsec:otherfindings}, we introduce silence thresholds on the gammatone spectrogram. Because NSIM is calculated on log-amplitudes, we found that it was too sensitive to different levels of ambient noise. For example, a near-digital silence reference compared against a very low level of ambient noise would still have a very low NSIM score, despite being perceptually transparent. The silence threshold introduces an absolute floor as well as a relative floor that may be higher for high amplitude frames.

The thresholded amplitude $y_{t,f}(x)$ for a time $t$ and frequency band $f$ given an input spectrogram $x$ is subject to:

\begin{equation}
y_{t,f}(x) = \max(Y_{min}, Y_{fmin}(t), x(t,f))
\end{equation}
where:
\begin{equation}
Y_{fmin}(t) = \max(r_{f}(t), d_{f}(t)) - Y_{fmin}
\end{equation}
given reference and degraded log amplitudes $r_{t,f}$ and $d_{t,f}$, and global absolute threshold $Y_{min}$, and relative per-frame threshold $Y_{fmin}$. 

\subsection{NSIM to MOS Model}
The changes above ultimately affect the NSIM scores. This requires that a new SVR model is trained to map the frequency band NSIM to MOS using libsvm~\cite{chang2011libsvm}. We conducted a grid search to minimize the 4-way cross validation loss on the same training set (TCDAudio14, CoreSV14, AACvOpus15). However, we observed in Section \ref{sec:case_studies} that this model was too specific to the training data and would behave poorly on very low bitrate (6-18~kbps) audio. This appears to be related to the fact that there is no monotonicity constraint in the SVR model used by ViSQOL (a strictly higher NSIM for out of distribution data produced lower MOS).  To address this issue for the default model, we relaxed the SVR parameters by lowering the cost and gamma parameters to have a slightly higher cross validation error while providing behavior that was closer to monotonic behavior.

Additionally, this version includes some tooling and documentation that allows for users to train their own SVR model by the use of CSV input files if the user can provide subjective scores for degraded/reference pairs. By following the grid search methods described by libsvm authors, users should be able to tailor a model that is able to represent their data.

\section{Discussion}
\label{sec:discussion}
Here we present a discussion of the use cases and feedback in light of the improvements. This is followed by reflection on trends and the areas that are promising as future work.

The case studies mentioned in Section \ref{sec:case_studies} highlight the challenges with real world applications of ViSQOL. The findings are generally that ViSQOL can be used for various applications, but careful investigation is required for any use case. The users of these tools are the very developers of new audio processing and coding techniques, and are often analyzing new types of audio that is ``out of distribution''. In some cases, we can allow the user to retrain a model to match the new data.

We find that developers are reasonably skeptical about how well ViSQOL will apply to their problem, given that it almost always has unique characteristics. Although ViSQOL is not guaranteed to give a meaningful absolute MOS for cases that are significantly different from what it was originally designed with, the developers in our case studies found some correlation that was useful for their use case.  However, this conclusion is often facilitated by the use of additional metrics that can be used to validate ViSQOL's application.

In other cases, for example, in the generative case, it is possible that it requires a redesign of the algorithm at a fundamental level, which could include different spectrogram representations or DNNs.  Projects like LibriTTS \cite{zen2019libritts} have curated large amounts of freely available speech data, which has been a boon to speech-related DNNs, there is yet no standard and widely available subjective score dataset that is of similar scale.  A larger dataset would enable new development, but also require rethinking of existing tools, such as support vector regression, used by ViSQOLAudio, which is intended for use on smaller datasets on the order of hundreds of points.
\section{Conclusion}
\label{sec:conclusion}
We have presented a new version of ViSQOL which is available for use on GitHub. The integration to real world problems by different teams at Google yielded a number of insights and improvements to the previous version. There are a number of promising avenues for future work, including DNN based approaches, a more general model, and taking the new generative audio approaches into account.

\section*{Acknowledgments}
We acknowledge Colm Sloan for his work on the initial C++ version of ViSQOL. This publication has emanated from research supported in part by the Google Chrome University Program and research grants from Science Foundation Ireland (SFI) co-funded under the European Regional Development Fund under Grant Number 13/RC/2289\_P2 and 13/RC/2077.

\bibliographystyle{IEEETran}
\bibliography{refs}

\end{document}